\documentclass[aps,pre,onecolumn,groupedaddress,showpacs]{revtex4}
\begin{document}

\title{The Rouse-Zimm-Brinkman theory of the dynamics of polymers in dilute solutions}%

\author{V. Lisy$^{a)}$} \author{J. Tothova}
\affiliation{Institute of Physics, P.J. Safarik University,\\ Jesenna 5, 041 54 Kosice, Slovakia}

\author{A.V. Zatovsky}
\affiliation{Department of Theoretical Physics, I.I. Mechnikov Odessa National University,\\
Dvoryanskaya 2, 65026 Odessa, Ukraine}
\date{\today}%

\begin{abstract}
We propose a theory of the dynamics of polymers in dilute solution, in which the popular Zimm and
Rouse models are just limiting cases of infinitely large and small draining parameter. The
equation of motion for the polymer segments (beads) is solved together with Brinkman's equation
for the solvent velocity that takes into account the presence of other polymer coils in the
solution. The equation for the polymer normal modes is obtained and the relevant time correlation
functions are found. A tendency to the time-dependent hydrodynamic screening is demonstrated on
the diffusion of the polymers as well as on the relaxation of their internal modes. With the
growing concentration of the coils in solution they both show a transition to the \textit{exactly}
Rouse behavior. The shear viscosity of the solution, the Huggins coefficient and other quantities
are calculated and shown to be notably different from the known results.

\end{abstract}


\maketitle

\section{Introduction}
Interest in dilute polymer solutions arises primarily from their importance in the
characterization of polymers, their interaction with solvent and from a fundamental interest in
understanding macromolecular response to hydrodynamic forces, free from the complications of
intermolecular entanglements. Despite five decades of investigations, these questions are not
entirely understood. So, there is a systematic discrepancy between the dynamic scattering data and
the theory \cite{Balabonov,Harnau,Dunweg}. The existing theories give different results for the
viscosity of dilute polymer solutions (see, \textit{e.g.}, \cite{Doi,Muthukumar} and the citations
there), the observed monomer motion in single polymer chains cannot be explained by the available
theories \cite{Shusterman}, the time dependence of the hydrodynamic screening in solutions is not
explained \cite{Ahlrichs}, \textit{etc}. For other problems we refer the reader to the recent
review \cite{Larson}. The aim of this work was to contribute to the solution of some of these
problems by developing a phenomenological bead-spring theory of the diffusion of an individual
"test" polymer and the relaxation of its internal modes in solution of unentangled polymers. Our
approach differs from the traditional ones in several main points. First, the joint Rouse-Zimm
theory is exploited \cite{Lisy}. In the literature, following de Gennes \cite{deGennes}, it is
assumed that at $\theta$ conditions the Zimm modes with the dispersion of relaxation times
$\tau_{pZ}\sim p^{-3/2}$ , where $p$ is the mode number, at low frequencies always must dominate
the Rouse modes with $\tau_{pR}\sim p^{-2}$; we show that both these contributions to the polymer
characteristics must be taken into account. Next, the internal modes are distributed discretely
(the assumption of their continuous distribution with respect to $p$ is true only in a restricted
time domain and often leads to incorrect interpretation of experimental data \cite{Tothova2005}).
The formalism from our theory with the time-dependent hydrodynamics of polymers has been adopted
\cite{Lisy,Tothova2003} and, finally, the Brinkman's theory \cite{Brinkman,Debye} for the flow in
porous media is used to take into account the influence of other coils on the studied test
polymer. The presented theory has the following limitations. The considered time scales are
$t>>\tau_R=R^2\rho/\eta$, where $R$ is the hydrodynamic radius of the polymer, $\rho$ is the
density and $\eta$ the viscosity of the solvent. This means that the effects of hydrodynamic
memory (or the viscous aftereffect) are neglected \cite{Lisy,Tothova2003,Zatovsky}. The
distribution of the coils in solution is considered to be stationary (this is justified at least
for the times $t<<\tau_D$; the choice of this time scale is possible since always
$\tau_p<<\tau_D$, where $\tau_D=R^2/D$ is the characteristic time of the coil diffusion with the
diffusion coefficient $D$). Our theory is also restricted to $\theta$ solvents \cite{Doi};
generalizations to other cases require knowledge of the equilibrium distribution of the segments
when the exclude volume interactions are taken into account. As already mentioned, only solutions
of unentangled polymers are considered. We are thus limited to concentrations of the chains
$c<1/[\eta]$, where $[\eta]$ is the intrinsic viscosity and $c$ the number of polymers per unit
volume \cite{Larson}. In spite of all these restrictions and other ones like those that we do not
consider the internal viscosity of polymers and the self-entanglements (the importance and even
the reality of these interactions are uncertain \cite{Larson}), we believe that the results of our
theory could be of interest. In particular, we have found new expressions for the quantities
describing the behavior of flexible polymers in solution, such as the diffusion coefficient of the
coil, the relaxation times of the internal modes, the viscosity of the solution, and the Huggins
coefficient. These quantities have been obtained from a generalized Rouse-Zimm equation for the
position vectors of the polymer segments and the Oseen tensor describing the velocity field of the
solvent due to perturbation. Finally, our theory describes in a simple manner the hydrodynamic
screening, \textit{i.e.} the concentration and time-dependent transition between the Zimm and (as
distinct from the previous theories) exact Rouse behavior of the polymer.

\section{The Rouse-Zimm-Brinkman theory of polymer dynamics}
We choose one coil as a "test" polymer. The equation of motion of its $n$th segment is
\begin{equation}\label{eq1}
M\frac{d^{2}\overrightarrow{x}_{n}(t)}{dt^{2}}=\overrightarrow{f}_{n}^{fr}
+\overrightarrow{f}_{n}^{ch}+\overrightarrow{f}_{n}.
\end{equation}
Here, $\overrightarrow{x}_n$ is the position vector of the segment (a spherical bead) from the $N$
ones constituting the polymer, $M$ is the bead mass, $\overrightarrow{f}_{n}^{ch}$ is the force
from the neighboring beads along the chain, $\overrightarrow{f}_{n}$ the random force due to the
motion of the molecules of solvent, and $\overrightarrow{f}_{n}^{fr}$ is the Stokes friction force
on the bead during its motion in the solvent \cite{Doi,Grosberg}:
\begin{equation}\label{eq2}
\overrightarrow{f}_{n}^{fr}=-\xi
\left[\frac{d\overrightarrow{x}_{n}}{dt}-\overrightarrow{v}\left(\overrightarrow{x}_{n}\right)\right],
\end{equation}
where $\overrightarrow{v}$ denotes the velocity of the solvent in the place of the $n$th bead due
to the motion of other beads. The friction coefficient on the bead with radius $b$ is
$\xi=6\pi\eta b$. This expression holds in the case of steady flow and takes into account the
hydrodynamic interaction. In a more general case with the hydrodynamic memory
\cite{Lisy,Tothova2003,Zatovsky} the force (2) should be replaced by the Boussinesq force and
equation (1) has to be solved together with the nonstationary hydrodynamic equations for the
macroscopic velocity of the solvent. To take into account the presence of other polymers in
solution, we use the Brinkman's work \cite{Brinkman} (see also \cite{Debye}) in which a polymer is
considered as a porous medium. In our approach all the solution is such a medium with coils being
obstacles to the solvent flow. Then in the right hand side of the Navier-Stokes equation a term
$-\kappa^2\eta\overrightarrow{v}$ has to be added, where $\kappa^{-2}$ is the solvent
permeability. This term has a sense of the average value of the force acting on the liquid in an
element of volume $dV$, provided the average number of polymers in solution per unit volume is
$c$; then $\kappa^2\eta=cf$, where $f$ is the friction factor on one coil. Thus, for an
incompressible solvent ($\nabla\overrightarrow{v}=0$)we have to solve the equation
\begin{equation}\label{eq3}
\rho \frac{\partial \overrightarrow{v}}{\partial t}=-\nabla p+\eta \triangle
\overrightarrow{v}-\kappa^{2}\eta\overrightarrow{v}+\overrightarrow{\varphi}.
\end{equation}
Here $p$ is the pressure and $\overrightarrow{\varphi}$ is the density of the force from the beads
of the studied polymer on the solvent \cite{Grosberg},
\begin{equation}\label{eq4}
\overrightarrow{\varphi}\left(\overrightarrow{x}\right)=-\sum_{n}\overrightarrow{f}_{n}^{fr}
\left(\overrightarrow{x}_{n}\right)\delta\left(\overrightarrow{x}-\overrightarrow{x}_{n}\right).
\end{equation}
To solve this equation is a difficult problem since the polymer chains are mobile. However,
restricting ourselves to the times much shorter than $\tau_D$, the concentration $c$ can be
assumed constant. The above equations then describe the motion of one bead in the solvent with the
effective influence of other coils on the motion of the solvent flow. This problem can be
transformed to that solved already in \cite{Zatovsky2001} (see also \cite{Zatovsky,Lisy}). The
velocity field can be in the Fourier representation in the time written as follows:
\begin{equation}\label{eq5}
v_{\alpha}^{\omega}\left(\overrightarrow{r}\right)=\int
d\overrightarrow{r}'\sum_{\beta}H_{\alpha\beta}^{\omega}\left(
\overrightarrow{r}-\overrightarrow{r}'\right)\varphi_{\beta}^{\omega}\left(\overrightarrow{r}'\right).
\end{equation}
Here the analog of the Oseen tensor is
\begin{equation}\label{eq6}
H_{\alpha\beta}^{\omega}\left(\overrightarrow{r}\right)=A\delta_{\alpha\beta}+Br_{\alpha}r_{\beta}r^{-2},
\end{equation}
\begin{eqnarray}\label{eq7}
A=(8\pi\eta r)^{-1}
\left\{e^{-y}-y\left[\left(1-e^{-y}\right)y^{-1}\right]''\right\},\nonumber\\
B=(8\pi\eta r)^{-1} \left\{e^{-y}+3y\left[\left(1-e^{-y}\right)y^{-1}\right]''\right\},
\end{eqnarray}
$y=r\chi$, $\chi^2=\kappa^2-i\omega\varrho/\eta$, and the prime means the differentiation with
respect to $y$. In the particular case $\omega= 0$ and for permeable solvent, $\kappa=0$, equation
(6) coincides with the well-known result of Zimm \cite{Doi}. Using this solution, a generalization
of the Rouse-Zimm equation can be obtained from the equation of motion \cite{Lisy}. The
preaveraging of the Oseen tensor over the equilibrium Gaussian distribution of the beads
\cite{Doi,Grosberg} gives
\begin{equation}\label{eq8}
\left\langle H_{\alpha \beta
nm}^{\omega}\right\rangle_{0}=\delta_{\alpha\beta}h^{\omega}\left(n-m\right),\overrightarrow{r}_{nm}\equiv\overrightarrow{x_{n}}-\overrightarrow{x_{m}},
\end{equation}
\begin{eqnarray}h^{\omega}\left(n-m\right)=\left(6\pi^3|n-m|\right)^{-1/2}(\eta a)^{-1}\left[1-\sqrt{\pi}z
\exp\left(z^2\right)\textrm{erfc}(z)\right].\nonumber
\end{eqnarray}
Here $a$ is the mean square distance between the beads along the chain and $z\equiv\chi a
\left(|n-m|/6\right)^{1/2}$. Then in the continuum approximation with respect to the variable $n$
the new Rouse-Zimm equation reads
\begin{equation}\label{eq9}
-i\omega\overrightarrow{x}^\omega(n)=\frac{1}{\xi}\left[\frac{3k_B T}{a^2}\frac{\partial^2
\overrightarrow{x}^\omega (n)}{\partial
n^2}+M\omega^2\overrightarrow{x}^\omega(n)+\overrightarrow{f}^\omega(n)\right]
\end{equation}
\begin{eqnarray}
+\int_{0}^{N} dm h^\omega(n-m)\left[\frac{3k_B T}{a^2}\frac{\partial^2\overrightarrow{x}^\omega
(m)}{\partial
m^2}+M\omega^2\overrightarrow{x}^\omega(m)+\overrightarrow{f}^\omega(m)\right].\nonumber\end{eqnarray}
It is solved with the help of the Fourier transformation (FT) in $n$, taking into account the
boundary conditions at the ends of the chain \cite{Grosberg}, $\partial
\overrightarrow{x}/\partial n=0$ at $n=0, N$: $\overrightarrow{x}^\omega(n)=\overrightarrow{y}_0
^\omega+2\sum_{p\geq 1}\overrightarrow{y}_{p}^{\omega}\cos(\pi np/N)$. The inverse FT then yields
\begin{equation}\label{eq10}
\overrightarrow{y}_{p}^{\omega}=\overrightarrow{f}_{p}^{\omega}\left[-i\omega
\Xi_{p}^{\omega}-M\omega^{2}+K_{p}\right]^{-1},
\end{equation}where
\begin{equation}\label{eq11}
\Xi_{p}^{\omega}=\xi\left[1+\left(2-\delta_{p0}\right)N\xi h_{pp}^{\omega}\right]^{-1},\\\
K_{p}= 3k_{B}T\left(\frac{\pi p}{N a}\right)^{2},\\\ p=0,1,2,...
\end{equation}
and the Oseen matrix is \cite{Lisy,Zatovsky}
\begin{equation}\label{eq12}
h_{pp}^\omega=\frac{1}{\pi\eta a \sqrt{3\pi N p}}\frac{1+\chi_p}{1+\left(1+\chi_p\right)^2},\\\chi_p\equiv\sqrt{\frac{N}{3\pi p}}\chi a,\\\
p=1,2,...,
\end{equation}
\begin{equation}\label{eq13}
h_{00}^\omega=\frac{2}{\sqrt{6N}\pi\eta a}
\frac{1}{\chi_\omega}\left[1-\frac{2}{\sqrt{\pi}\chi_\omega}+\frac{1}{\chi_\omega^2}
\left(1-\exp\chi_\omega^2\textrm{erfc}\chi_\omega\right)\right], \chi_\omega \equiv
\sqrt{\frac{N}{6}}\chi a.
\end{equation}
Using the fluctuation-dissipation theorem, the time correlation functions of the normal modes are
\begin{equation}\label{eq14}
\psi_p(t)=\left\langle y_{\alpha p}(0)y_{\alpha
p}(t)\right\rangle=\frac{k_{B}T}{\left(2-\delta_{p0}\right)\pi N}\int_{-\infty}^{\infty}d\omega
\cos\omega t\frac{
\textrm{Re}\Xi_{p}^{\omega}}{\left|-i\omega\Xi_{p}^{\omega}-M\omega^{2}+K_{p}\right|^{2}},
\end{equation}

\subsection{Diffusion of the coil}
In the stationary limit $\omega=0$ so that $\chi=\kappa$. Then the preaveraged Oseen tensor (6) is
\begin{equation}\label{eq15}
\left\langle H_{\alpha\beta}^\omega \right\rangle_0 =\left\langle\frac{\exp(-\chi
r)}{r}\right\rangle_0.\end{equation} The quantity $1/\kappa$ can be thus (for small $\kappa r$
only) considered as a screening length. For an individual polymer we had ($p=0$ in equation (14))
\cite{Lisy}
\begin{equation}\label{eq16}
\psi_{0}(0)-\psi_{0}(t)=D t
\end{equation}
with the diffusion coefficient $D=D_R+D_Z$ ($R$ and $Z$ stay for the well-known Rouse and Zimm
limits \cite{Doi}. Now instead of equation (12) we have $h_{00}^0$ with $\chi_0=\kappa R_G$ ($R_G$
being the gyration radius), the diffusion coefficient depends on the concentration of the coils
$c$,
\begin{equation}\label{eq17}
D=D_R+D_Z (c),
\end{equation}
($D_Z(0)=D_Z$) and consists of the Rouse (independent on the presence of other polymers) and the
Zimm contributions. The latter one can be expressed in the form
\begin{equation}\label{eq18}
D_Z (c)=D_Z f(c),
\end{equation}
where $f(c)$ is a "universal" function for every polymer:
\begin{equation}\label{eq19}
f(c)=\frac{3\sqrt{\pi}}{4\chi_0}\left[1-\frac{2}{\sqrt{\pi}\chi_0}+\frac{1}{\chi_0^2}
\left(1-\exp\chi_0^2\textrm{erfc}\chi_0\right)\right].
\end{equation}
The dependence of the permeability on the concentration is estimated as follows. The friction
coefficient in the quantity $\kappa^2=cf/\eta$ from equation (3) can be determined using the
Einstein relation $D=k_BT/f$. In such a picture
\begin{equation}\label{eq20}
\kappa^2=\frac{27\sqrt{\pi}}{16}\frac{\widetilde{c}}{R_G^2}\left(1+\frac{3}{4\sqrt{2}h}\right)^{-1}.
\end{equation}
Then the values of $\kappa$ and $\chi_0$ depend on the draining parameter $h=2(3N/\pi)^{1/2}b/a$
(if $h>>1$, the dynamics is of the Zimm type, for $h<<1$ we deal with the Rouse polymers). The
quantity $\widetilde{c}\equiv4\pi R_G^3c/3$ denotes the number of polymers per the volume of a
sphere with the radius $R_G$. With the increase of $c$ the Zimm term decreases and for large $c$
(small permeability $\kappa$ when $\chi_0>>1$) it becomes $\sim 1/\sqrt{c}$,
\begin{equation}\label{eq21}
D_Z(c)\approx\frac{2k_BT}{\pi\eta Na^2}\frac{1}{\kappa}.
\end{equation}
The realistic case of small $c$ corresponds to $\chi_0=\kappa R_G<<1$ when
\begin{equation}\label{eq22}
D_{Z}(c)=k_{B}T h_{00}^{0}(c)=D_{Z}\left(1-\frac{3}{8\sqrt{\pi}}\kappa R_{G}+...\right).
\end{equation}
The concentration dependent correction to $D_Z$ is thus proportional to $\sqrt{c}$ and differs
from other results (compare Ref. \cite{Zhao} and citations there, where this correction is $\sim
c$). The behavior of a free polymer depends on the draining parameter $h$. If $h$ is large, the
Zimm polymer (at $c=0$) with growing $c$ should change its behavior to the diffusion with the
Rouse coefficient $D_R$.

\subsection{Dynamics of internal modes}
In the stationary case ($\omega=0$) and at zero concentration ($\kappa=0$) the diagonal elements
of the Oseen matrix are well known \cite{Doi}. Now $h_{pp}^0$ from equation (13) depend on $c$.
The internal modes relax exponentially as in previous theories,
$\psi_p(t)\propto\exp(-|t|/\tau_p)$, but their relaxation rates consist of the Rouse contribution
and the concentration-dependent Zimm part,
\begin{equation}\label{eq23}
\frac{1}{\tau_{p}(c)}=\frac{1}{\tau_{pR}}+\frac{1}{\tau_{pZ}(c)},
\end{equation}
where $\tau_{pR}$ and $\tau_{pZ}(0)\equiv\tau_{pZ}$ are given in Refs. \cite{Doi,Grosberg} and
\begin{equation}\label{eq24}
\tau_{pZ}(c)=\frac{1}{2}\frac{1+\left(1+\chi_{p}\right)^{2}}{1+\chi_{p}}\tau_{pZ},
\end{equation}
which behaves as
\begin{equation}\label{eq25}
\tau_{pZ}(c)=\tau_{pZ}\left(1+\frac{N}{6\pi p}\kappa^2 a^2-...\right)
\end{equation}
if $c\rightarrow 0$, and (although unrealistic), as $c\rightarrow \infty$ one has
\begin{equation}\label{eq26}
\tau_{pZ}(c)\approx\frac{1}{2}\tau_{pZ}(0)\chi_{p}=\frac{\left(N a^{2}\right)^{2}\eta}{6\pi k_{B}T
p^{2} }\kappa.
\end{equation}
Note that for the internal modes the draining parameter depends on the mode number $p$:
$h(p)=\tau_{pR}/\tau_{pZ}=h/\sqrt{p}$. The "universal" dependence of $\tau_{pZ}(c)/\tau_{pZ}(0)$
on $\chi_p$ (24) indicates that with the growing $c$ every polymer shows a tendency to become the
Rouse one.

\subsection{Steady state viscosity and the Huggins coefficient}
The shear viscosity of the solution can be calculated from the formula \cite{Doi,Larson,Grosberg}
\begin{equation}\label{eq27}
\eta(c)=\eta+\frac{1}{2}k_{B}T c\sum_{p=1}^{\infty}\tau_{p}(c).
\end{equation}
Using equation (24), in the Rouse limit we have the familiar result \cite{Doi} $\eta(c)-\eta=\pi
N^2a^2bc\eta/6$. In the Zimm limit at small concentrations
\begin{eqnarray}
\frac{\eta(c)-\eta}{\eta}=\frac{c}{2\sqrt{3\pi}}\left(\sqrt{N}a\right)^{3}\zeta
\left(\frac{3}{2}\right)\left[1+c N a^{2}R_{Z}\zeta^{-1} \left(\frac{3}{2}\right)\zeta
\left(\frac{5}{2}\right)+...\right]
\end{eqnarray}
\begin{eqnarray}
=0.425c\left(N a^{2}\right)^{3/2}\left[1+0.140c\left(N a^{2}\right)^{3/2}+...\right]\nonumber,
\end{eqnarray}
where $R_Z$ is the Zimm hydrodynamic radius \cite{Doi} and $\zeta$ is the Riemann zeta function.
The first term coincides with the known result \cite{Doi}. A more general expression for the
viscosity, following from equations (27) and (24), is
\begin{equation}\label{eq29}
\frac{\eta(c)-\eta}{\eta}=\frac{1}{\pi}N^{2}a^{2}b
c\sum_{p=1}^{\infty}\frac{1}{p^{2}}\left(1+\frac{2h}{\sqrt{p}}\frac{1+\chi_{p}}{1+\left(1+\chi_{p}\right)^{2}}\right)^{-1}.
\end{equation}
At very low concentrations when $\chi_p<<1$ one has
\begin{equation}\label{eq30}
\frac{\eta(c)-\eta}{\eta}=\frac{1}{\pi}N^{2}a^{2}b
c\sum_{p=1}^{\infty}\frac{1}{p^{2}}\left(1+\frac{h}{\sqrt{p}}\right)^{-1}.
\end{equation}
Due to the dependence on $h$ the difference between this and the classical result \cite{Doi} can
be notable. So, for a polymer with small $h$ the ratio of the intrinsic viscosity
$[\eta]_h=\lim_{c\rightarrow 0} [\eta(c)-\eta]/(\eta c)$ at $h<1$ (when the polymer is assumed to
be the Rouse one) to that with $h=0$ changes as a function of $h$ from 1 to $\approx 0.55$, at
$h=0.5$ being 30 per cent smaller than in the case of a pure Rouse polymer. For a very large $h$
the intrinsic viscosity is $[\eta]_{h>>1}=3\sqrt{2/\pi}R_G^3\zeta(3/2)=6.253R_G^3$. Considering
the viscosity normalized to this expression, one can find that even for rather large $h$ the
difference from the traditional result for the pure Zimm polymer is significant. So, at $h=10$ it
represents some 25$\%$ and it is still about 10$\%$ even for $h$ such large as 50.

One of the important rheological parameters of polymer solutions is the Huggins coefficient $k_H$.
It can be determined from the general expression for the viscosity (29), using the intrinsic
viscosity $[\eta]_h$ at zero concentration (see equation (30)):
\begin{equation}\label{eq31}
\frac{\eta(c)-\eta}{\eta c}=[\eta]_h\left(1+k_H[\eta]_h c+...\right).
\end{equation}
We find
\begin{equation}\label{eq32}
k_H=\frac{3\pi}{2^{3/2}}\sum_{p=1}^\infty
\frac{1}{p^{7/2}}\left(1+\frac{h}{\sqrt{p}}\right)^{-2}\left(1+\frac{3}{4\sqrt{2}h}\right)^{-1}\left[\sum_{p=1}^\infty
\frac{1}{p^2}\left(1+\frac{h}{\sqrt{p}}\right)^{-1}\right]^{-2}.
\end{equation}
For large $h$ (the Zimm case) one thus has
\begin{equation}\label{eq33}
k_H=\frac{3\pi}{2^{3/2}}\zeta\left(\frac{5}{2}\right)\zeta^{-2}\left(\frac{3}{2}\right)\approx0.655.
\end{equation}
This value differs from the literature results, see, \textit{e.g.}, \cite{Doi} where $k_H=0.757$
is given; in Ref. \cite{Muthukumar1983} one finds $k_H=0.6949$, and in \cite{Freed1978} the
calculations gave the value 0.3787. The Freed and Edwards theory \cite{Freed1974,Freed1975}
possesses an intrinsic viscosity, which is inconsistent with the Kirkwood-Riseman steady-state
limit and gives the hydrodynamic screening even for infinitely dilute solutions (the discussion of
this question has been given already in the paper \cite{Muthukumar}).

As $h\rightarrow 0$ (the Rouse case), $k_H$ approaches zero as $k_H\approx 2\pi
h\zeta(7/2)\zeta^{-2}(2)$ and when $h$ grows, the Huggins coefficient slowly converges to the Zimm
limit (33). The difference from this limit is significant in a broad region of $h$, \textit{e.g.},
with the maximum $\approx 1.27$ of the function $k_H/k_{HZimm}$ at $h=3$, and with
$k_H/k_{HZimm}\approx 1.15$ for $h=20$.

\subsection{Monomer motion}
In connection with the unresolved problem of the dynamic nature of hydrodynamic screening in
polymer solutions (see Introduction), it is of special interest to consider the time-dependent
quantities describing the polymer behavior. Among such quantities, the relaxation modulus, which
determines the shear stress at shear flows can be easily studied since it is given simply by a sum
of exponentials containing the relaxation times from equation (23) \cite{Doi,Larson}. Here we
shall briefly focus on simplest (but observable \cite{Shusterman}) motion of the end monomer
within a polymer coil and calculate its mean square displacement (MSD). The MSD part due to
internal modes is \cite{Doi,Tothova2005}
\begin{equation}\label{eq34}
\left\langle r^2(t)\right\rangle_{int}=\frac{4Na^2}{\pi^2}\sum_{p=1}^\infty
\frac{1}{p^2}\left[1-\exp\left(-\frac{t}{\tau_p(c)}\right)\right]
\end{equation}
As already shown, with growing concentration $c$ every polymer tends to behave as a Rouse one,
which is due to the decrease of the Zimm contribution to the relaxation rates $\tau_p^{-1}$. The
time dependence of this screening is well displayed considering, \textit{e.g.}, the ratio of the
Rouse part of the MSD (\textit{i.e.} that if the polymer was the pure Rouse one, $h=0$) to the
total MSD in the joint Rouse-Zimm model. This function, $\langle r^2(t)\rangle_{int,R}/\langle
r^2(t)\rangle_{int}$, depends on the draining parameter $h$, the concentration $c$, and the time.
With the growing $t$ the above relation converges to unity showing the transition to the Rouse
behavior. For example, at a concentration $\widetilde{c}=0.1$ and $h=10$ we have $\langle
r^2(t)\rangle_{int,R}/\langle r^2(t)\rangle_{int}\approx0.75$ at $t=\tau_{1R}$, at $t=2\tau_{1R}$
the difference from the Rouse MSD is only about 10$\%$, and at $t=5\tau_{1R}$ the initially Zimm
polymer becomes indistinguishable from the Rouse one. When the same relation is considered as a
function of $\widetilde{c}$ for different times, one sees that the tendency to approach the Rouse
limit with the increase of $\widetilde{c}$ is more and more expressed as the time growths. At long
times, as expected, the polymer behaves as the Rouse one already at small concentrations.

\section{Conclusion}
The behavior of complex polymer systems that are attractive due to their unusual properties and
numerous applications cannot be understood without understanding the behavior of a single polymer
in a liquid, its interaction with the solvent and with other polymers in dilute solutions. Even in
situations when we deal with solutions of flexible and unentangled polymers, a number of open
questions exists for years and new "puzzles" appear. To our opinion, some of the problems are only
due to inappropriate use of the existing theories and to a great influence of simple "universal"
laws of polymer behavior, such as the famous $k^3$ law for the first cumulant of the dynamic
structure factor or the $t^\alpha$ laws for the monomer MSD, where $\alpha=1/2$ for the Rouse
polymer and 2/3 for the Zimm one. A closer look at these laws shows, however, that their
application to real situations is rather restricted and often they do not correspond to
experimental conditions. The model developed in the present work is not particularly new.
Partially (when dealing with the single polymer diffusion) it is known since the work by Kirkwood
and Riseman \cite{Kirkwood}. As to the internal polymer dynamics, our approach corresponds to that
by Dubois-Violette and de Gennes \cite{deGennes} who, however, have assumed that the internal
modes of the polymers should behave as the Zimm modes (\textit{i.e.} with the dispersion $\sim
p^{-3/2}$), thus neglecting the Rouse contribution, initially being present in their theory. Such
a simplification requires quantitative arguments and in many cases it is not substantiated, as
well as the assumption of the continuous distribution of the internal modes; we believe that it is
clearly shown in the present work. To take into account the presence of other coils in the
solution, we have used the well-known Debye and Bueche (or Brinkman's) theory for a porous medium.
Again, this approach has been already used in the polymer physics. However, coming from our
earlier results on the hydrodynamic theory of the polymer dynamics, we could in summary build a
model that is able to predict new results on the fundamental characteristics of polymer behavior
in dilute solutions. Some of the quantities characterizing the polymer solutions (viscosity,
Huggins coefficient) could be verified in standard experiments (with the necessary account for the
draining parameter). We have also proposed a description of the time dependence of the tendency to
hydrodynamic screening in dilute polymer solutions; this effect seems to be suitable for computer
simulations studies similar to those in Ref. \cite{Ahlrichs}.

\section*{Acknowledgment}
This work was supported by the Scientific Grant Agency of the Slovak Republic.

\end{document}